\documentclass[useAMS,usenatbib]{mn2e}
\usepackage{graphicx}

\title[Synchro-Curvature Self-Compton Radiation]{Synchro-Curvature Self-Compton Radiation of Electrons in Curved Magnetic Fields}
\author[Bo Zhang and Zi-Gao Dai]{Bo Zhang$^{1,2}$\thanks{zhangbo@nju.edu.cn}, Zi-Gao
Dai$^{1,2}$\thanks{dzg@nju.edu.cn}\\
$^{1}$Department of Astronomy, Nanjing University, Nanjing 210093,
P. R. China\\ $^2$Key laboratory of Modern Astronomy and
Astrophysics (Nanjing University), Ministry of Education, Nanjing
210093, P. R. China}

\begin{document}

\date{Accepted 00 00 00. Received 00 00 00; in original form 00 00 00}

\pagerange{\pageref{firstpage}--\pageref{lastpage}} \pubyear{2010}

\maketitle

\label{firstpage}

\begin{abstract}
In this paper we present the spectrum of synchro-curvature
self-Compton (SCSC) radiation of relativistic electrons with a
power-law distribution of Lorentz factors. We find that the
resulting spectrum is significantly different from that of either
synchrotron self-Compton or curvature self-Compton radiation if both
the curvature radius of the magnetic field and the cyclotron radius
of the electrons are within some proper ranges. The effects of
electrons' cooling and drifting, the low-energy self absorption in
seed spectra, and the Klein-Nishina cutoff are also discussed, in
order to get an accurate picture. We take gamma-ray bursts (GRBs) as
our example environment for discussions. The results would be
considered as a universal approach of the self-Compton emission of
relativistic electrons moving in curved magnetic fields, and thus
could be applied to many astrophysical phenomena, including GRBs,
active galactic nuclei (AGNs), and pulsars.
\end{abstract}


\section{Introduction}

Traditionally, non-thermal radiation mechanisms of electrons in
astrophysical environments are Bremsstrahlung, synchrotron
radiation, curvature radiation, and inverse Compton scattering (e.g.
see Blumenthal \& Gould 1970, and Rybicki \& Lightman 1979).
Bremsstrahlung radiation is the result of collisions between charged
particles, and has a continuous spectrum. Synchrotron emission
arises from relativistic electrons moving around straight magnetic
field lines, while curvature emission is radiated by electrons
moving along curved field lines, and can be used to discuss the
radiation from pulsar magnetosphere and AGNs (e.g., Cocke \&
Pacholczyk 1975). The formulae for calculating the characteristic
frequencies and spectral energy distributions of these two radiation
mechanisms are similar, while in the equations for curvature
radiation the curvature radius replaces the cyclotron radius in the
synchrotron radiation. In the high energy regime, the
inverse-Compton scattering usually plays an important role. This
type of emission occurs when relativistic electrons scatter low
energy photons.

For electrons with a power law distribution of Lorentz factors,
which are easily produced in astrophysical situations (e.g. products
of shock acceleration, see Fermi 1949 and Grupen et al. 2005), the
spectra of the resulting synchrotron and curvature radiations are
also in the form of power laws. The difference is that, due to
different relationships between Lorentz factor and critical
frequency, their spectral power law indices for the same index $p$
of electrons are not the same. The self-Compton emissions of these
two radiation mechanisms have spectral shapes similar to that of
seeds. Thus power law spectra, which are quite common under
astrophysical conditions, are explained. If seed photons are from
synchrotron radiation, which is the so-called synchrotron
self-Compton radiation, a similar spectrum to the synchrotron seed
is expected in this situation.

The synchro-curvature radiation depicting the spectrum emitted by
electrons moving around curved magnetic field lines was first
proposed by \citet{b15,b16,b2}, with a full set of formulae
depicting the radiation spectrum, radiation power, characteristic
frequency as well as polarization degree derived. The purpose of
considering this radiation mechanism is to give some insights into
new results which could not be interpreted well with conventional
mechanisms only. The related quantum radiation equations for a
single electron were given by \citet{b19}, and the spectra from
electrons with a power law energy distribution was calculated by
\citet{b20}. \citet{b7} gave further discussions on electrons with
larger transverse drifting velocities. As pointed out by
\citet{b10}, the synchro-curvature radiation should be considered as
a more realistic treatment to the radiations by electrons in the
universe.

Naturally, synchrotron and curvature radiations are the two limits
of synchro-curvature radiation: the curvature radius is infinite for
synchrotron radiation, while the electrons's cyclotron radius is
zero for curvature radiation. The synchro-curvature radiation can be
treated as a unified mechanism arising from relativistic electrons,
and thus can be applied to general researches related to
astrophysical radiations. It has already been used by to interpret
radiation theories of pulsars by \citet{b17,b18,b3,b8}, the energy
excess of active galactic nuclei by \citet{b27}, as well as spectral
observations of high energy photons from gamma-ray bursts observed
in the Compton Gamma-Ray Observatory (CGRO) Era by \citet{b4}. As
noted in \citet{b20}, due to the non-power law items in the formulae
describing the synchro-curvature radiation power, the spectrum is
significantly deviated from all of the ``traditional" mechanisms.
Besides, the polarization degrees of synchro-curvature, synchrotron
and curvature radiations are quite different from each other. So,
from observed spectral shapes and the polarization measurements, one
should distinguish between these three radiation mechanisms.

However, the synchro-curvature radiation alone cannot lead to a
complete picture. As noted in \citet{b20}, some other effects, such
as the Compton scattering of this radiation mechanism, needs to be
studied. Up to now synchro-curvature related inverse-Compton
scattering has remained undiscussed. In this paper we calculate the
spectrum of synchro-curvature self-Compton (SCSC) radiation, which
makes the scheme of synchro-curvature radiation more complete, and
present a more realistic approach to cosmic electron scattering. To
our knowledge, this is the first work on the SCSC radiation
mechanism. The structure of this paper is organized as follows. In
section 2 basic equations of both synchro-curvature radiation and
its self-Compton scattering are presented. In section 3 we show our
numerical calculation results with various parameters, along with
considerations of electron cooling, drifting and high energy
Klein-Nishina cutoff. We carry out these calculations mainly for the
case of gamma ray bursts, since these violent explosions are
representatives of high energy astrophysical events, and can provide
an extreme environment to discuss the high energy radiations. Our
results are summarized in section 4.

\section{Basic Equations}

In the following we consider the spectrum produced by relativistic
electrons moving in curved magnetic fields. The circular magnetic
field lines with constant curvature radius are assumed for
simplicity in all of our calculations. Of course this is just a
simplified treatment. In real situations, the curvature radius of a
magnetic field line may vary from place to place, and thus the
spectra will change accordingly. However, the basic equations for
calculation remain the same. All we need to do then is to replace
the curvature radius of field line with instant value.

First we present equations ignoring the electrons' drifts. As given
in \citet{b15,b16,b2} and \citet{b20}, the formula for the power per
frequency of a single electron moving around a magnetic field line
with curvature radius $\rho$ is written as

\begin{eqnarray}
\frac{{\rm{d}} P}{{\rm{d}} \omega} & = & \frac{{\rm{d}}
P_{\perp}}{{\rm{d}} \omega} + \frac{{\rm{d}} P_{\|}}{{\rm{d}}
\omega}
\nonumber\\
& = & \frac{\sqrt{3} e^2 \gamma \omega}{4 \pi
r_c^{*}\omega_c}\left\{ \left[ \int^{\infty}_{\omega/\omega_c}
K_{5/3}\left(y\right) {\rm{d}} y -
K_{2/3}\left(\frac{\omega}{\omega_c}\right) \right] \right.\nonumber\\
 & &+
\frac{\left[\left(r_B+\rho\right)\Omega_0^2 + r_B \omega_B^2
\right]^2}{c^4 Q_2^2}\left[\int^{\infty}_{\omega/\omega_c} K_{5/3}
\left(y\right){\rm{d}}y \right. \nonumber\\
& & \left. \left. + K_{2/3}\left(\frac{\omega}{\omega_c}\right)
\right] \right\},
\end{eqnarray}
and

\begin{eqnarray*}
\Omega_0 & = & \frac{c \cos \alpha}{\rho},\\
r_B & = & \frac{c\sin\alpha}{\omega_B}, \\
\omega_B & = & \frac{e B}{\gamma m_e c},\\
r_c^{*} & = & \frac{c^2}{\left(r_B + \rho\right) \Omega_0^2 + r_B \omega_B^2}, \\
\omega_c & = & \frac{3}{2} \gamma^3 c \frac{1}{\rho}\left[
\frac{\left(r^3_B + \rho r^2_B - 3 r_B \rho^2\right)}{\rho r_B^2}
\cos^4 \alpha + \frac{3 \rho }{r_B}\cos^2\alpha \right. \nonumber \\
& & \left.+\frac{\rho^2}{r_B^2}\sin^4 \alpha \right]^{1/2},\\
 Q_2^2 & = &
\left(\frac{r_B^2 + \rho r_B - 3 \rho^2}{\rho^3} \cos^3 \alpha \cos
\theta_0 + \frac{3}{\rho} \cos \alpha \cos \theta_0 \right.
\nonumber \\
& & \left.+ \frac{1}{r_B} \sin^3 \alpha \sin
\theta_0\right)\frac{1}{r_B},
\end{eqnarray*}
where $\Omega_0$ is the angular velocity of the moving electron's
guiding center, $r_B$ the cyclotron radius of the electron,
$\omega_B$ the cyclotron frequency, $r_c^{*}$ the instantaneous
curvature radius of the electron's trajectory, $\omega_c$ the
characteristic frequency of synchro-curvature radiation, and
$\alpha$ the angle between field lines and the electron's injecting
direction. Thus for power-law distributed relativistic electrons
with distribution index $p$, the emissivity $J_\nu$ could be
calculated according to following equations based on \citet{b20,b4}

\begin{equation}
J_{\nu} = J_{\perp}\left(\nu \right) + J_{\|}\left(\nu \right),
\end{equation}
and
\begin{eqnarray*}
J_{\perp}\left(\nu \right) & = &
\frac{\sqrt{3} e^2}{8\pi^2} \frac{9c^2}{4} \int N_0 \gamma^{-p+7}
\frac{1}{\nu_c^2}
\left\{ \frac{\nu}{\nu_c} \left[ \int^{\infty}_{\nu/\nu_c} K_{5/3}\left(y\right) {\rm{d}} y \right. \right. \nonumber\\
& & \left. \left.+ K_{2/3}\left(\frac{\nu}{\nu_c}\right) \right] \right\}\times\left(a_0 \gamma + a_1 + \frac{a_2}{\gamma}\right)^3 {\rm{d}} \gamma, \\
J_{\|}\left(\nu \right) & = & \frac{\sqrt{3} e^2}{2} \int N_0
\gamma^{-p+1}\left\{ \frac{\nu}{\nu_c} \left[
\int^{\infty}_{\nu/\nu_c} K_{5/3}\left(y\right) {\rm{d}} y \right.
\right. \nonumber\\
& & \left. \left.- K_{2/3}\left(\frac{\nu}{\nu_c}\right) \right]
\right\} \times \left( a_0 \gamma  + a_1 + \frac{a_2}{\gamma}
\right) {\rm{d}} \gamma,
\end{eqnarray*}
where we replace the circular frequency $\omega$ with angular
frequency $\nu = \omega/2\pi$ for convenience. The above two
equations give the emissivity in two polarizations, with $a_0 =
\frac{m_e c^2 \sin^2 \alpha \cos^2 \alpha}{e B \rho^2}$, $a_1 =
\frac{\cos^2 \alpha}{\rho}$, and $a_2 = \frac{eB \sin \alpha}{m_e
c^2}$. Therefore it is clearly seen that $a_0 \gamma + a_1 +
\frac{a_2}{\gamma} = \frac{1}{r^{*}_c}$.

While in real situations, electrons may not stay around one magnetic
field line; on the contrary, they can drift due to inhomogeneous
curved magnetic fields. This idea that charged particles can have
different motion in inhomogeneous magnetic field has already been
proposed several decades ago (e.g., see Chugunov et al. 1975), while
the unified formulae for single relativistic electron's
synchro-curvature radiation spectrum with electrons' transverse
drift velocity considered was given by \citet{b7} as follows:

\begin{eqnarray}
\frac{  { \rm{d} } P}{ { \rm{d} } \nu}& = & \frac{\sqrt{3} e^3}{m_e
c^2}    \left[ B^2 \sin^2 \alpha+ 2 \frac{\gamma m_e B}{e} \frac{
v_{\parallel}^2+ \frac{1}{2} v_{\perp}^2}{\rho} \sin \alpha \right.
\nonumber \\
 &  & \left.  + \frac{\gamma^2 m_e^2 }{e^2} \frac{\left(
 v_{\parallel}^2+ \frac{1}{2} v_{\perp}^2\right)^2}{\rho^2}\right]^{1/2}
 F\left(\frac{\nu}{\nu_c}\right),
\end{eqnarray}
where the characteristic frequency $\nu_c$ is defined by

\begin{eqnarray}
\nu_c & = & \frac{3 e }{2m_e c}  \left[  B^2 \sin^2 \alpha + 2
\frac{\gamma m_e B}{e} \frac{ v_{\parallel}^2+ \frac{1}{2}
v_{\perp}^2}{\rho} \sin \alpha \right. \nonumber\\
& & \left. + \frac{\gamma^2 m_e^2 }{e^2} \frac{\left(
 v_{\parallel}^2+ \frac{1}{2} v_{\perp}^2\right)^2}{\rho^2}
 \right]^{1/2}
 \frac{\gamma^2}{2 \pi},
\end{eqnarray}
where $F\left(x\right) = x \int^{\infty}_{x} K_{5/3}\left(y\right)
{\rm{d}} y$, $\alpha$ the angle between field lines and the
electron's injecting direction, $ v_{\parallel} = v \cos \alpha$ the
electron's velocity component parallel to the field, $v_{\perp} = v
\cos \alpha$ the electron's velocity perpendicular to the field, and
the electron's velocity $v$ can be calculated from the Lorentz
factor $\gamma$. It can be clearly seen that due to electrons'
drift, the spectral shape and flux are changed significantly.
Similar to the calculations for non-drifting electrons, the
emissivity of power law distributed electrons can be easily derived
from Eq.(3), with electrons' distribution function multiplied with
single electron's spectrum.

In Figure 1 we present the ratio between radiation powers at
characteristic frequency of synchro-curvature radiation without and
with the consideration of electrons' transverse drift velocity with
the various $B$ and $\rho$ used below. We fix the Lorentz factor as
$\gamma = 3 \times 10^3 $, which is typical for gamma ray bursts. It
can be seen that the relation between the ratio and the magnetic
field, especially between the ratio and $B$ is quite complex.

\begin{figure}
\begin{center}
\includegraphics[width=8cm]{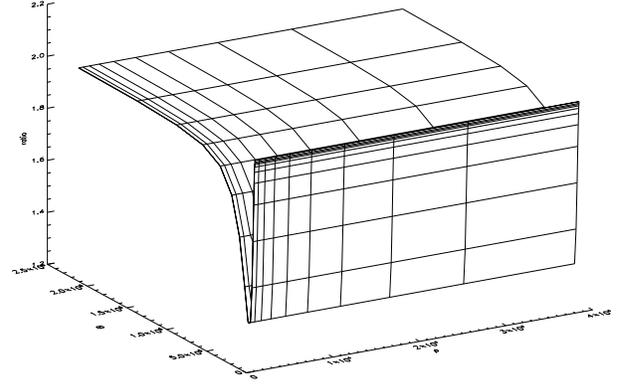}
\caption{\small{The ratio of synchro-curvature radiation's radiation
powers at characteristic frequency without and with the
consideration of electrons' transverse drift velocity in magnetic
fields with various parameters. X-axis denotes different values of
curvature radius $\rho$, while y-axis denotes different $B$.}}
\end{center}
\end{figure}

Similar to other traditional radiation mechanisms, the
synchro-curvature radiation may show self absorption features in the
low energy part in some situations. As noted in \citet{b25}, the
absorption coefficients of the two polarization components in vacuum
are

\begin{eqnarray}
K_{\nu, \perp} & = & \frac{p+2 }{4 \pi me \nu^2} \frac{\sqrt{3}
e^2}{2}\int N_0 \gamma^{-\left(p+1\right)} \frac{\gamma}{Q_2^2
r_c^{* 3} } \\ \nonumber & &\times \left[ \int^{\infty}_{\nu/\nu_c}
K_{5/3}\left(y\right) {\rm{d}} y + K_{2/3} \right] {\rm{d}} \gamma, \\
\nonumber K_{\nu, \|} & = & \frac{p+2 }{4 \pi me \nu^2}
\frac{\sqrt{3} e^2}{2}\int N_0 \gamma^{-\left(p+1\right)}
\frac{\gamma}{ r_c^{*} }\\ \nonumber & & \times \left[
\int^{\infty}_{\nu/\nu_c} K_{5/3}\left(y\right) {\rm{d}} y - K_{2/3}
\right] {\rm{d}} \gamma. \\ \nonumber
\end{eqnarray}

While in plasma environment, the coefficients may have some changes.
We take Eq.(2) and Eq.(3) as low energy seed spectra to calculate
the self-Compton spectrum of synchro-curvature radiation, and
consider the possible effects of self absorption. The IC emissivity
can be expressed as (e.g. Sari \& Esin 2001)

\begin{equation}
j_{\nu}^{IC} = 3 \sigma_T \int_{\gamma_m}^{\gamma_{max}} {\rm d}
\gamma N\left( \gamma \right) \int_0^{1} {\rm{d}} x
f_{\nu_s}^{\dagger} \left(x \right) g \left( x \right)
\end{equation}
with $x = \frac{\nu}{4 \gamma^2 \nu_s} $ and $ g \left( x \right) =
1+x+2x\ln x - 2x^2$. Here $f_{\nu_s}^{\dagger}$ means the incident
specific flux of frequency ${\nu_s}$ at the scattering region, and
can be related with the original seed flux $f_{\nu_s}$ by the
formula $f_{\nu_s} = f_{\nu_s}^{\dagger} \frac{4 \pi R^2}{4 \pi
D^2}$, with $R$ being the radius of the scattering region and $D$
being the distance to the observer. And $\nu$ is the frequency of IC
scattering, while $\nu_s$ is the frequency of the incident seed
emission.

The formulae above are only suitable in the Thomson limit. For the
high energy Klein-Nishina regime, we adopt the equation proposed by
\citet{b9} and quoted by \citet{b1} with modified scattering cross
section,

\begin{eqnarray}
j_{\nu}^{IC} & = & 3\sigma_T \int_{\gamma_m}^{\gamma_{max}} {\rm d}
\gamma N\left( \gamma \right) \int_{\nu_{s,min}^{\infty}}{\rm
d}\nu_{s}\frac{\nu f_{\nu_s}^{\dagger}} {4 \gamma^2 \nu^2_{s}}
\left[2y\ln y + y \right.\nonumber \\
& & \left.+1-2y^2 +\frac{1}{2} \frac{x^2 y^2}{1+xy}\left(1-y\right)
\right],
\end{eqnarray}
where $h$ denotes the Plank constant, $x = 4 \gamma h \nu_s
/\left(m_e c^2 \right)$, $y = h\nu/\left[x\left(\gamma m_e c^2 -h
\nu \right)\right]$, $\nu_{s,min} = \nu m_e c^2/\left[4\gamma\left(
\gamma m_e c^2 -h \nu \right)\right]$, and $\gamma > h\nu/\left(m_e
c^2\right)$.

Considering the polynomial items in the equations of the seed
spectrum, especially the ones that are $a_0\gamma + a_1 +
a_2/\gamma$ for non-drifting electrons and $\sqrt{B^2 \sin^2 \alpha
+ 2 \frac{\gamma m_e B}{e} \frac{ v_{\parallel}^2+ \frac{1}{2}
v_{\perp}^2}{\rho} \sin \alpha  + \frac{\gamma^2 m_e^2 }{e^2}
\frac{\left(v_{\parallel}^2+ \frac{1}{2}
v_{\perp}^2\right)^2}{\rho^2}}$ for drifting electrons, we need to
calculate the integrations above. The characteristic frequency of
SCSC is no longer a simple power law function of the electron
Lorentz factor. On the contrary, since different item dominates in
different band, the relationship between characteristic frequency
and Lorentz factor can be quite complex. \citet{b20} have already
shown that the seed synchro-curvature radiation spectra arising from
a power-law distributed electrons are quite different from simple
power law. Thus, a similar deviation is expected for SCSC.

\section{Numerical Results}

In the calculations below, we suppose that all of the electrons are
distributed isotropically for simplicity. In order to calculate the
modified Bessel functions of fractional order in the equations in
Section 2, we adopt the code provided by Chapter 6 of \citet{b11}.

In the calculations we mainly consider the situation of gamma-ray
bursts. GRBs are the strongest explosions in the universe since the
Big Bang, and the existence of strong magnetic fields as well as
relativistic electrons are implied both observationally and
theoretically (e.g., see Piran 1999). For this reason, such events
are often considered as laboratory to test some high energy process
in astrophysics. \citet{b4} applied the synchro-curvature radiation
to interpret the high energy excess observed by CGRO, and got
reasonable fitting results compared with other works using
traditional radiation mechanisms (e.g., see Tavani 1996). The
advantage of their work is that, by introducing the
synchro-curvature radiation, the number of free parameters can be
reduced, thus a more convenient treatment can be achieved. Here we
adopt the parameters similar to the fittings listed in \citet{b4}.
As noted in \citet{b21} and \citet{b22}, typically speaking, the
bulk Lorentz factor of GRBs is $\Gamma \sim 10^2 - 10^3$, magnetic
field $B \sim 10^4$, and electrons' minimum Lorentz factor
$\gamma_{min} \sim 10^2 - 10^3$. The fitting results provided by
\citet{b4} are around $\Gamma\left[\gamma_{min}^2 B\right]_{12} \sim
10$ and $\Gamma\left[\gamma_{min}^3 /\rho \right]_{8} \sim 10$,
where $\left[\gamma_{min}^2 B\right]_{12} = \gamma_{min}^2
B/10^{12}$ and $\left[\gamma_{min}^3 /\rho \right]_{8} =
\left(\gamma_{min}^3 /\rho \right)/10^{8}$.

These results yield relatively small curvature radius of magnetic
field. In theory, several magnetic field configuration models have
been proposed for GRBs. As noted in \citet{b28}, one of the most
popular models is that, the magnetic field in the GRB jets can be
produced by the shock, and are compressed into a thin layer normal
to the propagation direction. Since the field in such a layer is
highly tangled, small scale variations of magnetic field (therefore
small value for field's characteristic curvature radius $\rho$) can
be achieved. Some numerical simulations on GRB magnetic field
involving the two stream (Weibel) instability also give rise to
similar long-lasting field structure, especially in internal shocks.
(e.g. see Medvedev \& Loeb 1999, Medvedev et al. 2005 and references
herein). And according to \citet{b37} and \citet{b38}, small scale
structures of fields can dominate in the case of GRBs. In such
situations, it is possible that synchro-curvature radiation as well
as SCSC can play an important role in shaping the GRBs' spectrum.

First we show the seed spectra samples in Figure 2. The solid line,
dotted line and dashed line correspond to the spectrum of
synchro-curvature radiation, synchrotron radiation, and curvature
radiation respectively.

\begin{figure}
\begin{center}
\includegraphics[width=8cm]{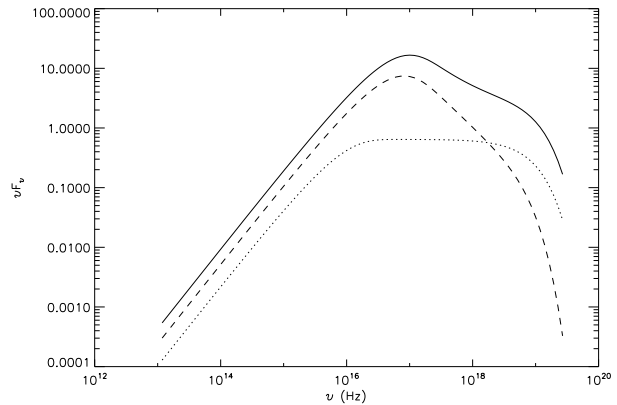}
\caption{\small{Spectra of synchro-curvature radiation (solid line),
synchrotron radiation (dashed line), and curvature radiation (dotted
line) with magnetic field $B = 10^4$ Gauss, curvature radius of
field $\mathbf{B}$ $\rho = 10^3$ cm, minimum electron Lorentz factor
$\gamma_{min} = 10^3$, maximum Lorentz factor $\gamma_{max} = 10^4$,
and electron distribution index $p=5$. Fluxes are in arbitrary
unit.}}
\end{center}
\end{figure}

From Figure 2, we can see that for synchrotron radiation rising from
electrons with power law distribution of Lorentz factors, the
spectral index should be $\left(p-1\right)/2$ (where $p$ is the
power law index of electrons). While for curvature radiation, the
index is $\left(p-2\right)/3$. However, due to polynomial items in
the synchro-curvature radiation formulae, the resulting spectrum
cannot be described with a single power law. Thus a turnoff is
expected for self-Compton emission, and corresponding to a point
where a different item in the polynome dominates. Roughly speaking,
this is the point when $a_0 \gamma  =  a_2/\gamma$. Figure 3 shows
the corresponding IC spectrum of Figure 2.

\begin{figure}
\begin{center}
\includegraphics*[width=8cm]{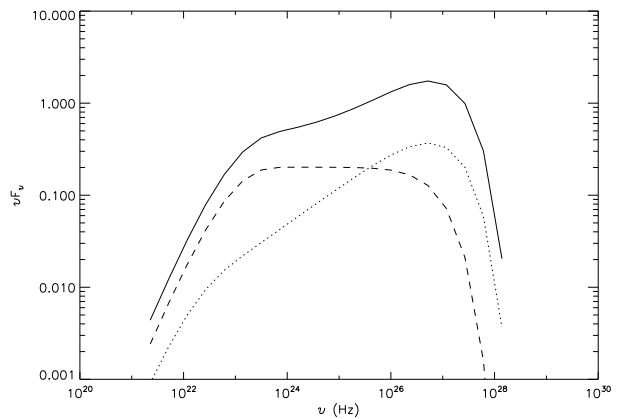}
\caption{\small{Self Compton spectra of synchro-curvature radiation
(solid line), synchrotron radiation (dashed line), and curvature
radiation (dotted line). Parameters are the same as in Figure 1.}}
\end{center}
\end{figure}

Figures 2 and 3 show the spectra of electrons with a large power law
index ($p=5$). For shock acceleration which is common in
astrophysical circumstances, electrons usually have a smaller index,
$p \sim 2-2.5$. We present calculational results for such electrons
in Figure 4.

\begin{figure}
\begin{center}
\includegraphics*[width=8cm]{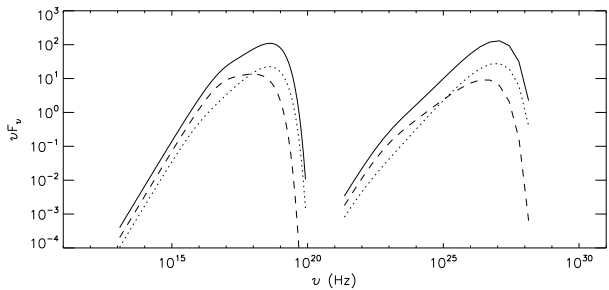}
\caption{\small{Synchro-curvature radiation (solid line),
synchrotron radiation (dashed line), and curvature radiation (dotted
line) spectra (left) and corresponding self Compton spectra (right)
for electron collective with $p=2.5$. Other parameters are the same
as in Figure 1.}}
\end{center}
\end{figure}

It is noted that for a smaller index $p$, the spectrum doesn't show
a significant turnoff. We try to find some explanations for this
phenomena. In the seed spectrum, synchrotron and curvature
radiations have different spectral index $\alpha$. Although the
synchro-curvature emission is not a simple addition to the former
two traditional radiation spectra, we can see that the
synchro-curvature spectrum can be considered as a summation of
synchrotron and curvature plus the coupling items. If electrons can
have cyclotron radius value similar to magnetic field's curvature
radius (which mean the contributions from synchrotron and curvature
radiations are comparable with each other), as well as large
difference between spectral indices for synchrotron and curvature
radiations, the high energy turnoff can be very evident. For
parameters within the ranges used in our calculation, it is seen
from Figures 2, 3 and 4 that for a smaller curvature radius, weaker
magnetic fields and larger electron index $p$, synchrotron and
curvature radiations overlay more significantly. Since the resulting
self Compton emissions have similar (although smoother) spectral
shapes to seeds, turnoff points are expected naturally.

The turnoffs can also be explained by the equations describing the
resulting spectrum. The ratio of the $a_0 \gamma$ to $a_2/\gamma$
items of the polynomy $\left(a_0\gamma + a_1 + a_2/\gamma\right)$
can change a lot across the observed energies. So the spectral index
may change from one frequency range to another. Comparing their
relative values, we can get an overall idea of the position of the
turnoff point. Figure 2 from \citet{b20} gives an example of the
relationship between the magnetic field $B$ and the energy of
turnoff point. For self-Compton scattering a similar relationship is
expected.

Besides, if we change $\gamma_{max}$ to a larger value, the shape of
the high energy turnoff can also be affected. Usually the larger
$\gamma_{max}$, the longer the excess part. The exact value of
$\gamma_{max}$ is related to the particle acceleration and radiation
mechanisms, e.g., in the case of gamma ray bursts, electrons whose
shock acceleration timescale equals to cooling timescale of
non-thermal radiation have the maximum Lorentz factor (e.g, see Dai
\& Lu 1998). Thus the value $\gamma_{max}$ can be calculated. And
the values of $B$, $\rho$ and so on also affect the final shape of
the spectra: smaller $B$ and $\rho$ lead to a more significant
spectral deviation from traditional IC scattering, while spectra
formed in magnetic fields with larger $B$ and larger curvature
radius $\rho$ are nearly identical with SSC radiation.

Similar to other types of inverse Compton scattering, the
polarization of SCSC depends on the magnetic field structure, the
distribution of emitting photons, as well as viewing angle.
Anisotropic electrons can give rise to significant polarized IC
emission (e.g. see Begelman \& Sikora 1987). As noted by
\citet{b36}, usually in previous works related to GRBs, isotropic
electrons are assumed (e.g., see Granot et al. 1999), although this
may not be the real situation. With random fields, the seed spectra
from the whole jet do not have large degrees of polarization; the
polarized signals from different regions are wiped out. Thus one can
not expect too much degrees of polarization for SCSC emission in the
comoving frame. Since the polarization degree remains the same with
Lorentz transformation, the observed emission for on-axis observer
does not show significant polarization either. However anisotropy
emerges to the off-axis observers. Thus polarization patterns for
SCSC exist in this situation, and are differs from traditional
radiation mechanisms, especially in the high energy regime, since
the polarization can be affected by the spectral shape. Besides, it
is possible that non-zero polarization degree arise from the
underlying field with some degree of order, that is, the field is
not totally random. And some works also show that relativistic
electrons do have some anisotropy in the comoving frame (e.g., see
Achterberg et al. 2001). Some recent works on GRBs have taken
anisotropy of electrons into account (e.g., see Beloborodov et al.
2010), although this issue need to be discussed in more details.
Therefore more detailed simulations may be needed to describe the
real polarization behavior.

Finally we consider the self absorption effects of the
synchro-curvature seed spectrum. Since the absorption only show
significant effects in the low energy regime, changing the spectra
index, while the differences between SCSC and other traditional
mechanisms are mainly in the high energy portion (that is, the high
energy turnoff), the self absorption effects do not matter a lot in
our discussion. So we do not consider this issue in the following
calculations for simplicity. In fact, for synchro-curvature seed
radiation in vacuum, the major absorption part is quite similar to
synchrotron self absorption, with the spectrum proportional to
$\nu^{5/2}$, and only a bit flatter in the higher energy part of the
absorption regime, as noted in \citet{b25}. However, negative self
absorption may occur in plasma environment, although still affect
the low energy part only.

\subsection{Cooling Effect of Electrons in the Case of Gamma-Ray Bursts}

Integration over the electron Lorentz factor $\gamma$ is needed to
get the whole spectrum. In the calculation of the spectra shown in
Figures 2 to 4, a simple power law distribution is assumed. However,
under real astrophysical circumstances, the cooling effect of
electrons should be considered in order to get an accurate spectrum,
e.g., in the situation to calculate the spectrum of gamma-ray bursts
(e.g., see Sari, Piran \& Narayan 1998). Here we take this issue
into account, and discuss the fast and slow cooling regimes for GRBs
under the framework of SCSC.

Let $p$ denote the initial power law index. For fast cooling
electrons, all of the electrons can cool down to about the cooling
Lorentz factor (at which the cooling timescale equals to the
dynamical timescale of the system). In this situation, the
distribution function should be

\begin{equation}
N\left(\gamma \right) = \left\{
\begin{array}{cc}
n_{\gamma} \gamma^{-2}, \gamma_c < \gamma <\gamma_m \\
n_{\gamma} \gamma^{p-1}_m \gamma^{-p-1}, \gamma_m < \gamma < \gamma_{max},\\
\end{array}
\right.
\end{equation}
where $n_{\gamma} = \gamma_c n_e$, $n_e$ is the number of electrons,
$\gamma_m$ the minimal Lorentz factor, $\gamma_c$ the cooling
Lorentz factor, and $\gamma_{max}$ the maximum Lorentz factor. In
the fast cooling regime, $\gamma_c < \gamma_m < \gamma_{\max}$.
Under such conditions, the spectra of synchro-curvature emission and
its self Compton radiation are drawn in Figures 5 and 6.

\begin{figure}
\begin{center}
\includegraphics*[width=8cm]{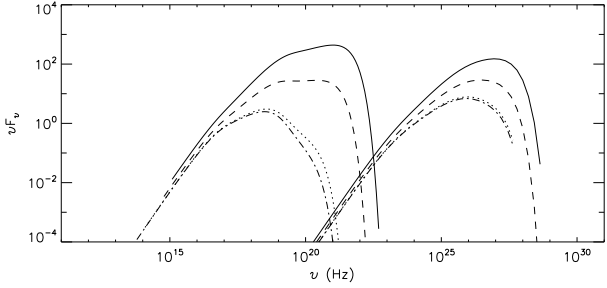}
\caption{\small{Curves on the left show the synchro-curvature
radiation spectra of fast cooling electrons with index $p=5$,
magnetic field $B = 10^4 $ Gauss and curvature radius $\rho = 0.5
\times 10^3$ cm (solid line), $\rho = 1 \times 10^3$ cm (dashed
line), $\rho = 5 \times 10^3$ cm (dotted line), and $\rho = 10
\times 10^3$ cm (dashed dotted line). The corresponding self Compton
spectra are shown on the right. All of the spectra are in arbitrary
units.}}
\end{center}
\end{figure}

\begin{figure}
\begin{center}
\includegraphics*[width=8cm]{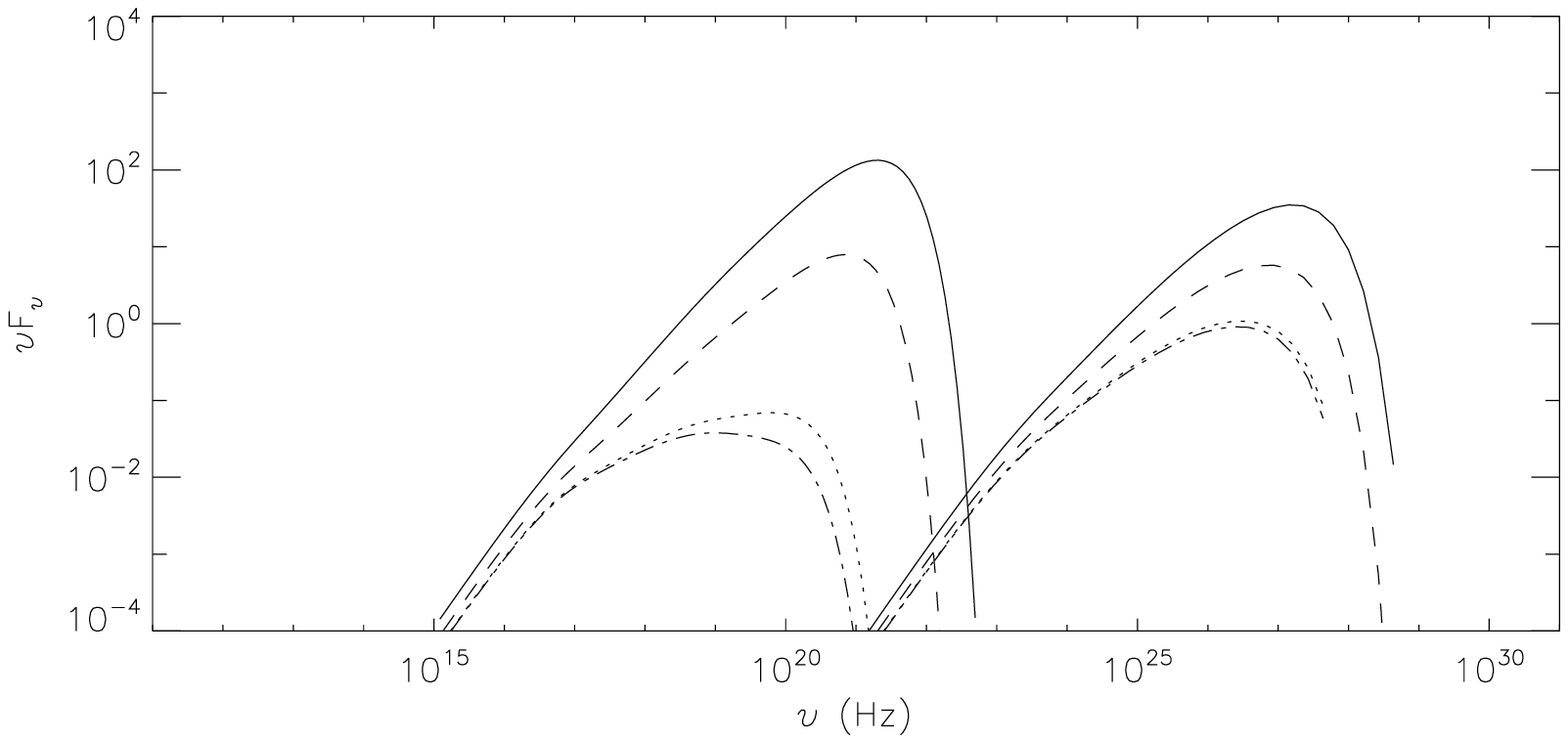}
\caption{\small{Curves on the left show the synchro-curvature
radiation spectra of fast cooling electrons with index $p=2.5$,
magnetic field $B = 10^4 $ Gauss and curvature radius $\rho = 0.5
\times 10^3$ cm (solid line), $\rho = 1 \times 10^3$ cm (dashed
line), $\rho = 5 \times 10^3$ cm (dotted line), and $\rho = 10
\times 10^3$ cm (dashed dotted line). The corresponding self Compton
spectra are shown on the right. All of the spectra are in arbitrary
units.}}
\end{center}
\end{figure}

For slow cooling electrons, $\gamma_m < \gamma_c < \gamma_{\max}$,
and $n_{\gamma} = \left(p-1 \right)\gamma_m^{p-1} n_e$. The electron
distribution is

\begin{equation}
N\left(\gamma \right) = \left\{
\begin{array}{cc}
n_{\gamma} \gamma^{-p}, \gamma_m < \gamma <\gamma_c \\
n_{\gamma} \gamma_c \gamma^{-p-1}, \gamma_c < \gamma < \gamma_{max}.\\
\end{array}
\right.
\end{equation}

\begin{figure}
\begin{center}
\includegraphics*[width=8cm]{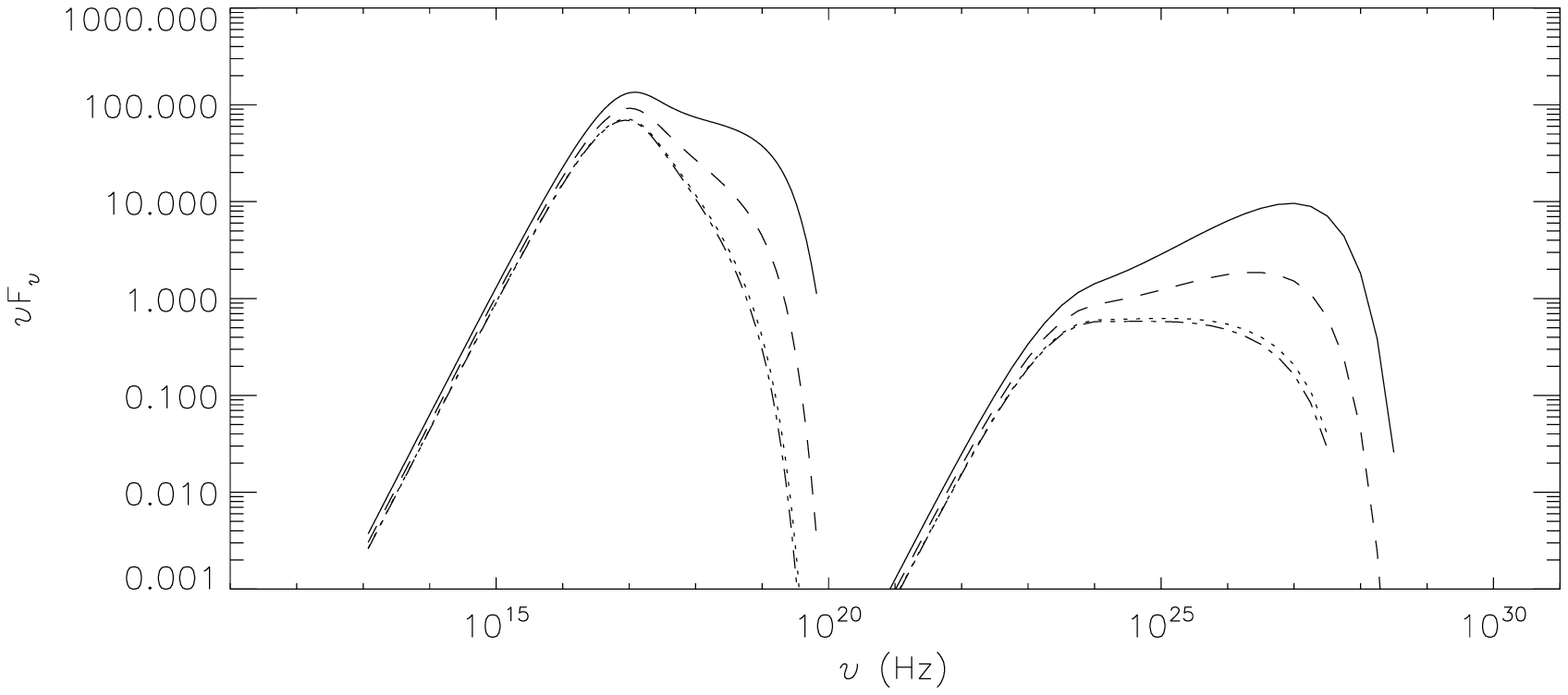}
\caption{\small{Curves on the left show the synchro-curvature
radiation spectra of slow cooling electrons with index $p=5$,
magnetic field $B = 10^4 $ Gauss and curvature radius $\rho = 0.5
\times 10^3$ cm (solid line), $\rho = 1 \times 10^3$ cm (dashed
line), $\rho = 5 \times 10^3$ cm (dotted line), and $\rho = 10
\times 10^3$ cm (dashed dotted line) as IC scattering seeds. The
corresponding self Compton spectra are on the right. All of the
spectra are in arbitrary units.}}
\end{center}
\end{figure}

\begin{figure}
\begin{center}
\includegraphics*[width=8cm]{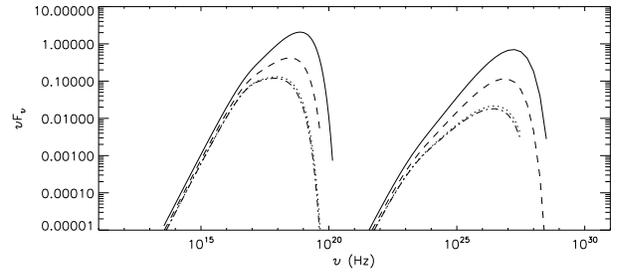}
\caption{\small{Curves on the left show the synchro-curvature
radiation spectra of slow cooling electrons with index $p=2.5$,
magnetic field $B = 10^4 $ Gauss and curvature radius $\rho = 0.5
\times 10^3$ cm (solid line), $\rho = 1 \times 10^3$ cm (dashed
line), $\rho = 5 \times 10^3$ cm (dotted line), and $\rho = 10
\times 10^3$ cm (dashed dotted line) as IC scattering seeds. The
corresponding self Compton spectra are shown on the right. All
spectra are in arbitrary units.}}
\end{center}
\end{figure}

In Figures 7 and 8, sample spectra of slow cooling electrons are
presented. It is clearly seen that the electron cooling effect makes
the low energy index of electrons smaller, thus making the resulting
spectrum nearly identical with synchrotron self-Compton radiation,
especially for fast cooling electrons that have a smaller low energy
index. For slow cooling electrons with a larger low energy index,
the difference between SSC and SCSC becomes more significant.

For other astrophysical objects, e.g., the cooling effects can be
different. \citet{b33} has given a detailed view of this issue.

\subsection{Spectra of Drifting Electrons}

As shown in Section 2, the resulting synchro-curvature radiation can
be changed if electron drifts exist due to inhomogeneous magnetic
field. The formulae describing drifting electrons' spectra are quite
different from the ones emitted by non-drifting electrons. Thus the
self Compton emission from such electrons is also not the same as
simple SCSC shown above. Here we calculate drifting electrons'
synchro-curvature self Compton radiation.

\begin{figure}
\begin{center}
\includegraphics*[width=8cm]{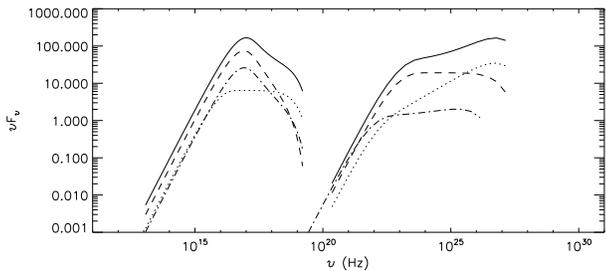}
\caption{\small{Spectra on the left are from the synchro-curvature
radiation without drifts (solid line), the synchro-curvature
radiation with drifts (dash-dotted line), the synchrotron radiation
(dashed line), and the curvature radiation (dotted line) with $B =
0.5 \times 10^4$ Gauss, $p=5$ and $\rho = 10^3$ cm. The
corresponding self Compton spectra are shown on the right. All of
the spectra are in arbitrary units.}}
\end{center}
\end{figure}

In Figure 9 we show the seed and self Compton spectra of
synchro-curvature radiation with and without drifts, as well as the
synchrotron and curvature radiations. Because of the difference in
the distribution of seed photons, the drifting effects can change
the resulting SCSC spectra significantly, especially in the high
energy regime.

For non-drifting electrons, the high energy regime in the SCSC
spectra is mainly contributed by the curvature self Compton
radiation. The large transverse velocity can suppress the electrons'
motion along magnetic field lines and thus suppress the curvature
radiation part. Thus such deviation can be explained.

\subsection{Klein-Nishina Corrected Spectrum}

All of the spectra presented above are calculated in the Thomson
limit. For high energy photons, i.e., $\gamma h \nu > m_e c^2$,
where $\nu$ denotes the frequency, and $\gamma$ the Lorentz factor
of electrons, the scattering cross section does not remain constant
as the Thomson cross section $\sigma_T$. In this situation, the
Klein-Nishina cross section formula should be applied instead of
$\sigma_T$. Some Klein-Nishina corrected synchro-curvature spectrum
samples with various parameters are presented in Figures 10 and 11.

\begin{figure}
\begin{center}
\includegraphics*[width=8cm]{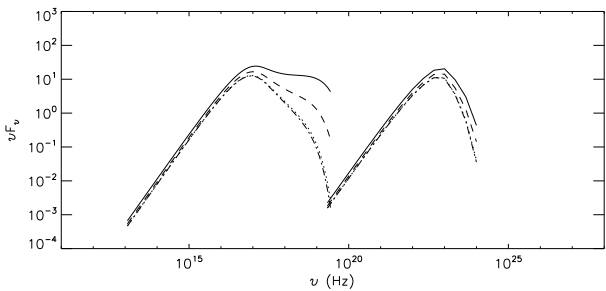}
\caption{\small{Spectra on the left are from the synchro-curvature
radiation spectra of electrons with index $p=5$ and curvature radius
$\rho = 0.5 \times 10^3$ cm (solid line), $\rho = 1 \times 10^3$ cm
(dashed line), $\rho = 5 \times 10^3$ cm (dotted line), and $\rho =
10 \times 10^3$ cm (dashed dotted line). The corresponding self
Compton spectra with KN cross section correction are shown on the
right. All of the spectra are in arbitrary units.}}
\end{center}
\end{figure}

\begin{figure}
\begin{center}
\includegraphics*[width=8cm]{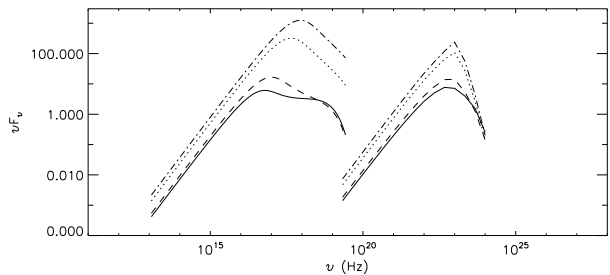}
\caption{\small{Spectra on the left are from the synchro-curvature
radiation spectra with different magnetic fields. The solid line
shows spectrum with $B = 0.5 \times 10^4$ Gauss, the dashed line $B
= 1 \times 10^4$ Gauss, the dotted line $B = 5 \times 10^4$ Gauss,
and the dashed dotted line $B = 10 \times 10^4$ Gauss. The
corresponding KN corrected self Compton spectra (right) are for
electron collective with $p=5$. Other parameters are the same as in
Figure 10.}}
\end{center}
\end{figure}

From Figures 10 and 11 it is clearly seen that due to the high
energy suppression of the Klein-Nishina scattering cross section,
the high energy turnoff of SCSC radiation in the Thomson limit no
longer exists. So we conclude that the difference between SCSC and
SSC could not be significant when the Klein-Nishina cutoff is
dominated.

\section{Summary and Discussions}

In this paper we present the spectrum of the SCSC radiation for
various parameters and situations. We here summarize all of the
results listed above:

First of all, the real radiation mechanism for relativistic
electrons in magnetic fields could not be the synchrotron or
curvature radiation. On the contrary, the synchro-curvature
radiation provides a more realistic treatment. Thus a similar
situation is expected for self Compton scattering. If the seed
spectrum is synchro-curvature rather than other traditional
mechanisms, the inverse-Compton spectrum could still be an open
question. In fact, it can be seen from the contents above that the
resulting spectra deviates from the form of simple power law,
especially for electrons with a larger distribution index and the
drifts of electrons in magnetic fields are ignored. A curved
magnetic field can significantly changes the resulting spectrum.
Thus SSC alone can not precisely describe the real IC spectra.
Combined with the synchro-curvature seed, multiple turnoffs in the
whole spectrum could be expected.

Second, the resulting inverse-Compton spectrum of synchro-curvature
radiation can be affected by various parameters, including the
strength of the magnetic field $B$, the curvature radius of the
field $\rho$, the index of the electron energy distribution $p$, and
so on. For example, a smaller curvature radius $\rho$ and a smaller
field strength $B$ usually leads to a more significant high energy
turn off. The resulting spectra from electrons with a smaller
distribution index resembles to synchrotron self Compton radiation,
while for electrons with a larger index of the Lorentz factor
distribution, the self-Compton spectrum of synchro-curvature
radiation clearly shows a high energy turnoff in the Thomson regime.
Besides, drifting effects of electrons due to inhomogeneous magnetic
field can also change the SCSC spectra, with different shapes of
turnoffs. This phenomena can be attributed to the different
relations between the characteristic frequency and the radius of
electrons' motion for synchrotron and curvature radiations (the SCSC
spectrum is something like the summation of the synchrotron and
curvature self-Compton radiations, although not exactly), as well as
the polynomial items in the formulae describing the spectrum. So the
high energy self-Compton spectra of a certain astrophysical source
may also used as a possible probe of seed radiation mechanisms, and
the spatial and temporal variations of radiation spectra can reveal
the changes in magnetic field.

Third, the cooling process of electrons can change the electron
distribution index. As a result, the resulting scattering spectra is
also changed. As noted above, the high energy turnoff of SCSC
radiation only becomes significant when the electron distribution
index is large. Fast cooling electrons have a relatively small low
energy index, and they could suffer more influences. So the high
energy turnoff may be not as significant as expected in real
situations.

Fourth, it should be noticed that the calculations in Section 3 are
all based upon the assumption of isotropically distributed
electrons. We calculated the electrons with $\delta$-shape
distribution function as well, and the resulting spectra are quite
similar to the ones from isotropic electrons. However, the spectra
might be slightly different from the sample results presented in
this paper for electrons with other forms of spatial distribution.

Fifth, the self absorption (positive for vacuum, both positive and
negative are possible for plasma) may occur in the synchro-curvature
seed spectrum. However, this absorption is only evident in the low
energy part, thus nearly have no effect on the high energy regime.
Since the SCSC radiation differs from traditional radiation
mechanisms mostly in the high energy band, self absorption is not a
major concern for the calculated SCSC radiation spectra and
distinguishing SCSC from other mechanisms.

Sixth, since we mainly discuss the self Compton radiation for
isotropically distributed electrons from GRB shock acceleration, the
polarization in the seed spectra can be nearly wiped out by random
electrons. So the polarization degree is nearly zero in the
scattering spectrum for on-axis observers. This may not be the real
situation, and high polarization degree may exist for off-axis
observations, highly anisotropic electrons, or magnetic fields with
a certain preferred direction. While the exact value of the degree
depends on the configuration of electrons' distribution, and careful
simulations may be required to calculate this.

Finally, in the Klein-Nishina regime, the difference between SCSC
and SSC radiation as well as other traditional forms of
inverse-Compton scattering is relatively small due to strong high
energy suppression. Although the difference could not be ignored in
the Thomson regime, in the Klein-Nishina limit it seems difficult to
distinguish between the synchro-curvature self Compton radiation
from traditional inverse-Compton scattering.

The detection of synchro-curvature radiation as well as SCSC can
provide vital clues of the structure of magnetic field. Here we also
take GRB as example. The field configuration in the jet may be
randomly distributed (as noted in Section 3) or ordered field (e.g.,
see Toma et al. 2009 and references herein). One way to distinguish
these two models is to measure the polarization in the GRB
emissions. However, current instruments are not quite suitable for
this task; no reliable polarization measurement on high energy
emission has been obtained yet. However, if one can analyze the
spectrum and find out any synchro-curvature or SCSC component, this
could lead to a conclusion that random field may be the real
configuration; since ordered magnetic field have nearly straight
field lines, and thus can not produce synchro-curvature radiation.
On the other hand, if both of the spectral signature and
polarization degree can be measured in the future, the understanding
on GRB magnetic field will be greatly improved.

As noted at the beginning of Section 2, in this paper only circular
field lines are considered. While in real astrophysical
environments, the curvature radius of magnetic fields usually
changes from place to place, as one can expected. Thus these
calculations only show the emission from a certain area at a certain
time, and analyzing the spectra's spacial behavior (for extended
sources) and temporal evolution can provide some insights into the
structure of the magnetic fields at the source region. For example,
if one can detect the transition between synchro-curvature/SCSC
radiation and other traditional radiation mechanisms in a single
source, the magnetic field distribution can be unveiled, at least in
some degrees.

Similar to the synchro-curvature seed spectrum, the self-Compton
spectrum presented in this paper should be considered as the
baseline of IC scattering calculations, and can be used universally
to explain the high energy observations (especially the ``abnormal''
spectral behavior, including the high energy excess), and can be
applied to nearly all high energy astronomical objects, e.g., active
galactic nuclei, gamma-ray bursts, and pulsars. For example, as
noted in \citet{b4}, several GRBs observed in the CGRO era have high
energy turnoffs; and similar spectral features has also been
detected by Fermi Gamma Ray Space Telescope, with possible high
energy exponential cutoff observed in one burst (GRB 090926A, e.g.,
see Ackermann et al. 2010). The SCSC provides another possible
explanation for these facts, since SCSC spectra yield high energy
turnoff, while limited maximum Lorentz factor can naturally give
rise to high energy cutoff. And KN effects of SCSC can provide sharp
cutoff as well. Here multiband data is required in order to tell the
real radiation mechanism beneath. And probing the real mechanism
through high energy observations is also possible, especially for a
large energy index of electrons, although in the Klein-Nishina
regime things can be a bit more difficult.

Of course, the high energy turnoff in observational data could be
easily explained by the superimposition of several spectra arising
from different mechanisms, so it's a complicated task to confirm the
contributions from synchro-curvature radiation and its self Compton
scattering, especially when the scattering occurs in the
Klein-Nishina limit with strong high energy suppression (thus the
overall spectrum may not show multiple turnoffs). In order to solve
this problem, temporal data are required. If both the low energy
part and the high energy turnoff evolve simultaneously, SCSC
radiation is a possible radiation mechanism; or several mechanisms
will take the responsibility, since different mechanisms usually
show different temporal behaviors. However exceptions may also exist
in the latter case, since the structure change of a magnetic field
can also lead to different spectra at different times.

\section*{Acknowledgments}

The authors thank Bing Zhang for helpful discussions and an
anonymous referee for valuable comments that have allowed us to
improve the manuscript. This work is supported by the National
Natural Science Foundation of China (grant no. 10873009 and
11033002) and the National Basic Research Program of China (973
program) No. 2007CB815404.

\bsp

\label{lastpage}

\end{document}